\documentclass[%
 reprint,
superscriptaddress,
 amsmath,amssymb,
prb,
]{revtex4-1}
\usepackage{amsmath}
\usepackage{natbib}

\usepackage{graphicx}
\usepackage{soul}
\usepackage{mhchem, textcomp}
\usepackage{graphicx}
\usepackage{dcolumn}
\usepackage{bm}
\usepackage{wrapfig}
\usepackage{graphicx, type1cm, lettrine}
\usepackage{float}
\restylefloat{figure}
\usepackage{hyperref}

\begin{document}

\preprint{APS/123-QED}

\title{Oxide-mediated self-limiting recovery of field effect mobility in plasma-treated MoS$_2$}

\author{Jakub Jadwiszczak}
\affiliation{School of Physics, Trinity College Dublin, Dublin 2, Ireland}
\affiliation{Centre for Research on Adaptive Nanostructures and Nanodevices (CRANN), Trinity College Dublin, Dublin 2, Ireland} 
\affiliation{Advanced Materials and BioEngineering Research Centre (AMBER), Trinity College Dublin, Dublin 2, Ireland}

\author{Colin O'Callaghan}
\affiliation{School of Physics, Trinity College Dublin, Dublin 2, Ireland}
\affiliation{Centre for Research on Adaptive Nanostructures and Nanodevices (CRANN), Trinity College Dublin, Dublin 2, Ireland} 
\affiliation{Advanced Materials and BioEngineering Research Centre (AMBER), Trinity College Dublin, Dublin 2, Ireland}

\author{Yangbo Zhou}
\affiliation{School of Physics, Trinity College Dublin, Dublin 2, Ireland}
\affiliation{Centre for Research on Adaptive Nanostructures and Nanodevices (CRANN), Trinity College Dublin, Dublin 2, Ireland} 
\affiliation{Advanced Materials and BioEngineering Research Centre (AMBER), Trinity College Dublin, Dublin 2, Ireland}
\affiliation{School of Material Science and Engineering, Nanchang University, 999 Xuefu Road, Nanchang, Jiangxi, China, 330031}

\author{\\Daniel S. Fox}
\affiliation{School of Physics, Trinity College Dublin, Dublin 2, Ireland}
\affiliation{Centre for Research on Adaptive Nanostructures and Nanodevices (CRANN), Trinity College Dublin, Dublin 2, Ireland} 
\affiliation{Advanced Materials and BioEngineering Research Centre (AMBER), Trinity College Dublin, Dublin 2, Ireland}

\author{Eamonn Weitz}
\affiliation{School of Physics, Trinity College Dublin, Dublin 2, Ireland}

\author{Darragh Keane}
\affiliation{Centre for Research on Adaptive Nanostructures and Nanodevices (CRANN), Trinity College Dublin, Dublin 2, Ireland} 
\affiliation{Advanced Materials and BioEngineering Research Centre (AMBER), Trinity College Dublin, Dublin 2, Ireland}
\affiliation{School of Chemistry, Trinity College Dublin, Dublin 2, Ireland}

\author{Ian O'Reilly}
\affiliation{School of Physics, Trinity College Dublin, Dublin 2, Ireland}

\author{\\Clive Downing}
\affiliation{Centre for Research on Adaptive Nanostructures and Nanodevices (CRANN), Trinity College Dublin, Dublin 2, Ireland} 
\affiliation{Advanced Materials and BioEngineering Research Centre (AMBER), Trinity College Dublin, Dublin 2, Ireland}

\author{Aleksey Shmeliov}
\affiliation{Centre for Research on Adaptive Nanostructures and Nanodevices (CRANN), Trinity College Dublin, Dublin 2, Ireland} 
\affiliation{Advanced Materials and BioEngineering Research Centre (AMBER), Trinity College Dublin, Dublin 2, Ireland}
\affiliation{School of Chemistry, Trinity College Dublin, Dublin 2, Ireland}

\author{Pierce Maguire}
\affiliation{School of Physics, Trinity College Dublin, Dublin 2, Ireland}
\affiliation{Centre for Research on Adaptive Nanostructures and Nanodevices (CRANN), Trinity College Dublin, Dublin 2, Ireland} 
\affiliation{Advanced Materials and BioEngineering Research Centre (AMBER), Trinity College Dublin, Dublin 2, Ireland}

\author{John J. Gough}
\affiliation{School of Physics, Trinity College Dublin, Dublin 2, Ireland}
\affiliation{Centre for Research on Adaptive Nanostructures and Nanodevices (CRANN), Trinity College Dublin, Dublin 2, Ireland} 

\author{Cormac McGuinness}
\affiliation{School of Physics, Trinity College Dublin, Dublin 2, Ireland}

\author{\\Mauro S. Ferreira}
\affiliation{School of Physics, Trinity College Dublin, Dublin 2, Ireland}
\affiliation{Centre for Research on Adaptive Nanostructures and Nanodevices (CRANN), Trinity College Dublin, Dublin 2, Ireland} 
\affiliation{Advanced Materials and BioEngineering Research Centre (AMBER), Trinity College Dublin, Dublin 2, Ireland}

\author{A. Louise Bradley}
\affiliation{School of Physics, Trinity College Dublin, Dublin 2, Ireland}
\affiliation{Centre for Research on Adaptive Nanostructures and Nanodevices (CRANN), Trinity College Dublin, Dublin 2, Ireland} 

\author{John J. Boland}
\affiliation{Centre for Research on Adaptive Nanostructures and Nanodevices (CRANN), Trinity College Dublin, Dublin 2, Ireland} 
\affiliation{Advanced Materials and BioEngineering Research Centre (AMBER), Trinity College Dublin, Dublin 2, Ireland}
\affiliation{School of Chemistry, Trinity College Dublin, Dublin 2, Ireland}

\author{Valeria Nicolosi}
\affiliation{Centre for Research on Adaptive Nanostructures and Nanodevices (CRANN), Trinity College Dublin, Dublin 2, Ireland} 
\affiliation{Advanced Materials and BioEngineering Research Centre (AMBER), Trinity College Dublin, Dublin 2, Ireland}
\affiliation{School of Chemistry, Trinity College Dublin, Dublin 2, Ireland}

\author{Hongzhou Zhang}
\email{hozhang@tcd.ie}
\affiliation{School of Physics, Trinity College Dublin, Dublin 2, Ireland}
\affiliation{Centre for Research on Adaptive Nanostructures and Nanodevices (CRANN), Trinity College Dublin, Dublin 2, Ireland} 
\affiliation{Advanced Materials and BioEngineering Research Centre (AMBER), Trinity College Dublin, Dublin 2, Ireland}

\begin{abstract}

Precise tunability of electronic properties of 2D nanomaterials is a key goal of current research in this field of materials science. Chemical modification of layered transition metal dichalcogenides leads to the creation of heterostructures of low-dimensional variants of these materials. In particular, the effect of oxygen-containing plasma treatment on molybdenum disulfide (MoS$_2$) has long been thought to be detrimental to the electrical performance of the material. Here we show that the mobility and conductivity of MoS$_2$ can be precisely controlled and improved by systematic exposure to oxygen:argon plasma, and characterise the material utilising advanced spectroscopy and microscopy. Through complementary theoretical modelling which confirms conductivity enhancement, we uncover the role of a two-dimensional phase of molybdenum trioxide (2D-MoO$_3$) in improving the electronic behaviour of the material. Deduction of the beneficial role of MoO$_3$ will serve to open the field to new approaches with regard to the tunability of 2D semiconductors by their low-dimensional oxides in nano-modified heterostructures.

\end{abstract}

\maketitle 


\lettrine[lines=3]{T}{he} recent decade has produced intense research into layered two-dimensional nanomaterials,
with transition metal dichalcogenides (TMDs) such as MoS$_2$ being the prime focus in the area of novel nanoelectronics \cite{Radisavljevic_2011,Radisavljevic2013a,Jariwala_2014,Lembke2015,Sangwan_2015}. Progress demands that a nanofabrication
methodology is developed to control the structure and properties of semiconducting layered crystals
so that desired functionalities are obtained for these materials in the future. These may include phase transitions \cite{Lin_2014,Kim_2016} or conductivity modulation for next-generation data storage \cite{Fox_2015,Cheng_2016}. \newline
\indent In particular, the interaction of low energy RF-generated
plasma ions with MoS$_2$ has already led to the creation of a multitude of
devices, including rectifying diodes, photovoltaics and
non-volatile memories \cite{Chen_2013, Wi_2014, Chen_2014}. Plasma power and exposure time have emerged as key variables to delineate between chemical etching and physical sputtering regimes \cite{Cui_1999, Fleischauer_1987,Park_2005,Kim_2016b}. Treatment with oxygen-containing plasma leads to the formation
of molybdenum trioxide (MoO$_3$) centres which have been reported to increase the resistivity of the
material and inhibit carrier transport, while retaining relative
structural integrity of the now oxide-containing MoS$_2$ heterostructure \cite{Islam_2014,Khondaker_2016}. 
Here we demonstrate the tuning of electrical resistivity of few-layer
MoS$_2$ by treatment with O$_2$:Ar (1:3) plasma. The field effect mobility, $\mu_{\ce{FE}}$, of the MoS$_2$ channel is seen to deteriorate initially but recovers to above-original levels after 6 seconds of exposure to the plasma. The associated electrical conductivity of the devices is noted to increase by an order of magnitude at this stage. Upon further treatment, the conductivity and mobility drop again and no subsequent recovery is seen. \newline
\indent In the limited literature regarding this phenomenon, the reason for the apparent recovery remains under debate \cite{Chen_2016,Yang_2015,Nan_2017}.
In this article, we propose a mechanism of impurity-mediated electrical tuning facilitated by a two-dimensional phase of MoO$_3$, with advanced spectroscopic and microscopic studies to support electrical characterisation. We infer the presence of this 2D molybdenum trioxide phase, which serves to provide additional charge carriers to the MoS$_2$ channel at 6 seconds of plasma treatment which increases its conductivity. Complementary mathematical modelling of conductive networks reveals the beneficial effect of the freshly-incorporated oxide in the MoS$_2$ matrix. \newline
\indent Recent theoretical work has predicted the two-dimensional phase
of MoO$_3$ to be a material with a distinctly high acoustic
phonon-limited carrier mobility ($>$ 3000 cm$^{2}$ V$^{-1}$ s$^{-1}$)
\cite{Zhang_2017}, while experimental 2D FETs made of sub-stoichiometric exfoliated MoO$_3$ have reported mobilities far exceeding that of MoS$_2$ ( $>$
1100  cm$^{2}$ V$^{-1}$ s$^{-1}$) \cite{balendhran2013enhanced,alsaif2016high}. The advantageous effect of the 2D phase of MoO$_3$ on the electrical properties of MoS$_2$ may play a key role in the applications of planar heterostructures of layered TMDs in novel electronic devices. Future research into this area must consider the benefits of defect-mediated transport in 2D nanoelectronics. \\ \\
\noindent \textbf{Recovery of field effect mobility in plasma-treated MoS$_2$} \newline
\noindent For the initial plasma exposures, the level of drain-source current for a 4 layer (4L) device varies slightly until 6 seconds, when a significant rise in output current is noted indicating an increase in the conductivity of the 
\onecolumngrid

{\setlength\intextsep{5pt}
\begin{figure}[H]
\centering
\includegraphics[scale=1]{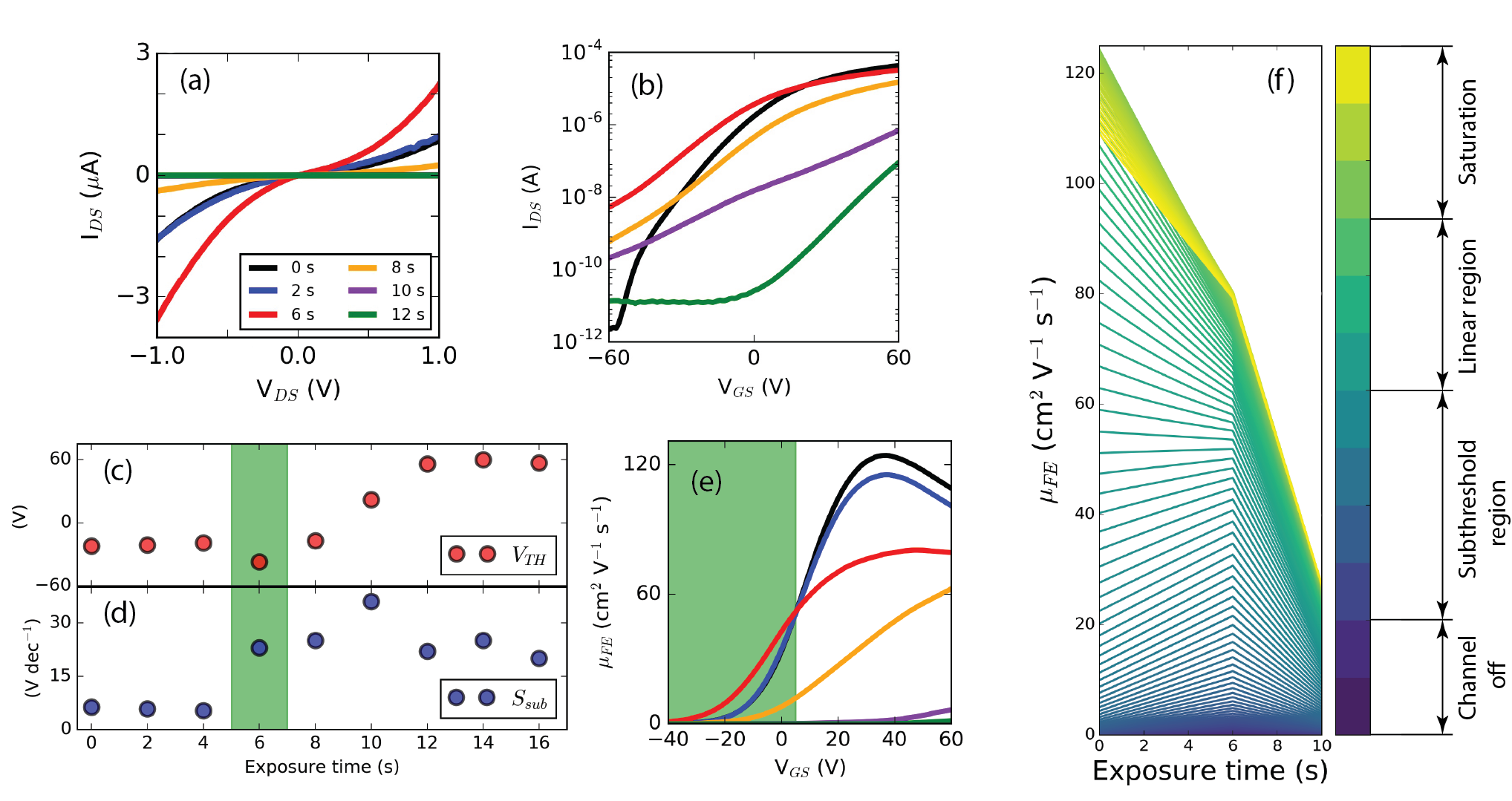}
\caption{{\footnotesize \textbf{Note that (a),(b) $\&$ (e) share the same color legend. (a) I-V curve evolution over exposure time for a 4L device. All curves are measured at zero gate bias. We note the increased current density observed after 6 seconds (red). (b) Gate sweeps of same device over exposure time. The curves after 6 s, 8 s and 10 s show a largely linear response in the semi-log plot at low gate biases and do not reach threshold over standard sweep range. Subsequent treatment until 12 s drastically shifts V$_{\ce{TH}}$ to positive gate biases and lowers the current by several orders of magnitude. (c) Threshold voltage for the same device shows a sudden drop at 6 seconds followed by a steady increase to extremely positive gate biases over treatment time. (d) Subthreshold swing variation with exposure time shows a diminished response to the gate field starting at 6 seconds. The area marked in green in (c) $\&$ (d) indicates the electrical recovery region. (e) Extraction of field effect mobility for the same 4L device across the whole gate bias range (graph begins from - 40 V for clarity). The peak mobility reached in the curves is shown to degrade over time. The green region marks the area of the gate sweep where the 6 s exposure attains highest relative mobility values (red curve). (f) Mobility change over plasma treatment time extracted at gate biases between - 60 V and 60 V. The color legend explicitly maps the curves onto different regions of the gate sweep. The increase at 6 seconds is visualised in the form of rising recovery peaks in the body of the plot, which correspond to 6 s mobilities evaluated in the green area marked in (e).}}}
\label{fig:electrical}
\end{figure}}
\twocolumngrid 
\noindent channel (\hyperref[fig:electrical]{\textbf{Fig.1a}}). Subsequent exposures cause a continuing drop in current level until the noise floor of the instrument (10$^{-11}$ A) is reached past 12 seconds of plasma treatment. The gate characteristics (\hyperref[fig:electrical]{\textbf{Fig.1b}}) of the n-type MoS$_2$ change significantly after 6 seconds. The level of output current at negative gate values increases by several orders of magnitude at 6 s, implying a drastic shift in the threshold voltage (V$_{\ce{TH}}$) to negative gate biases. Correlated with this is the change in the sensitivity of the output current to the applied gate-source bias (V$_{\ce{GS}}$), with a much more gradual increase in output current throughout the sweep. \hyperref[fig:electrical]{\textbf{Figures 1c,d}} track the evolution of the threshold gate voltage (V$_{\ce{TH}}$) and the subthreshold swing ($\ce{S_{sub}}$) over plasma exposure time. The threshold voltage is seen to shift from $\sim$ -21 V for the untreated device to $\sim$ -37 V at 6 s of exposure and subsequently to large positive gate biases of $\sim$ 22 V at 10 s and $\sim$ 60 V at $>$ 10 s. The shift towards negative threshold voltages at 6 s implies increased depletion mode functionality for n-type devices, while the upshift of V$_{\ce{TH}}$ after further exposure denotes an increase of p-type doping. $\ce{S_{sub}}$, in turn, initially shows little change until it increases 5-fold at 6 s and up to 8-fold at 10 s relative to the values before treatment. Upon further exposure, it drops again to $\sim$ 25 V dec$^{-1}$ as V$_{\ce{TH}}$ is shifted to large positive gate biases. At 6 seconds, all samples show a marked increase in $\ce{S_{sub}}$, indicating that they are less sensitive to variations in gate field around the region where the FET conductive channel is formed. \newline
\indent The field effect mobility, $\mu_{\ce{FE}}$, of the device is plotted as a function of gate bias in \hyperref[fig:electrical]{\textbf{Fig.1e}}. The peak value of mobility is seen to drop with exposure time, and the gate bias necessary to reach saturation shifts towards larger V$_{\ce{GS}}$. Importantly, for V$_{\ce{GS}}$ in the region from -60 V to 5 V (highlighted green area in the plot), the mobility at 6 s can now be tuned to much higher values than for the untreated device, with corresponding conductance of the device increasing by over one order of magnitude in this region. Subsequent treatments to 8 s and 10 s decrease $\mu_{\ce{FE}}$ markedly and no recovery is seen beyond this point.  \newline
\indent In addition, the change in $\mu_{\ce{FE}}$ over exposure time at each applied gate bias between -60 V and 60 V is charted in \hyperref[fig:electrical]{\textbf{Fig.1f}} in 1 V steps. The extracted curves are color-mapped to the palette seen on the right, scaling from extremely low gate biases (-60 V) to extremely high (60 V). Inside the decay envelope of the peak mobility evident from the edge contour of this graph, we observe a series of recovery peaks around the region corresponding to a treatment time of 6 seconds. This recovery is pronounced in the linear regime near V$_{\ce{TH}}$, i.e. where $\mu_{\ce{FE}}$ rises above initial values extracted for the untreated device. This corresponds directly to the green region in \hyperref[fig:electrical]{\textbf{Fig.1e}} where the red curve (6 s) attains higher values than the other curves, i.e. across V$_{\ce{GS}} \in$ [-60 V, 5 V]. The subsequent drop in $\mu_{\ce{FE}}$ and conductance is a direct consequence of material etching and introduction of scattering centres that happens after 10 seconds. \newline
\indent The notable increase in current density seen in \hyperref[fig:electrical]{\textbf{Fig.1a}}, the close-to-linear response to the variation in gate bias at 6, 8 and 10 s (red, purple and orange curves in \hyperref[fig:electrical]{\textbf{Fig.1b}}), and the increase in $\ce{S_{sub}}$ all hint at the presence of a highly conductive phase in the material at 6 seconds of exposure, which is responsible for the recovery. In the following sections, we show this phase is a two-dimensional form of MoO$_3$ produced by a chemical reaction with the plasma. \\

\noindent \textbf{MoS$_2$ surface modification by oxygen insertion} \newline
\noindent We use atomic force microscopy (AFM) to track the thickness variation and surface roughness of the plasma-treated flake. Phase maps of the same region on a 4L flake are shown in \hyperref[fig:structural]{\textbf{Figs.2a-c}}, with notable change in contrast indicating material difference over time. \hyperref[fig:structural]{\textbf{Fig.2d}} charts the change in the edge heights evaluated from line profiles across the edges of 4L and 5L regions (see \textbf{Supplementary Fig. 9}). The initial edge height on the 4L portion increases by $\sim$ 30$\%$ from 0 s to 6 s, and on the 5L area by $\sim$ 10$\%$. This is followed by a subsequent drop in height for longer exposure times. The surface roughness (\hyperref[fig:structural]{\textbf{Fig.2e}}) stays constant within instrument precision, and does not vary more than the thickness of one layer of MoS$_2$ or MoO$_3$ in the first 8 seconds of exposure. \newline
\indent The peak in edge height at 6 s is a critical point at which the etching mechanism shifts from one largely dominated by chemical oxygen insertion into the lattice, to one where argon-dominated sputtering of material and removal of species from the surface takes over. With increasing dose of the plasma, the integration of oxygen into the mechanically-exfoliated 2H-MoS$_2$ structure will introduce considerable change; including rearrangement of electronic density and effective lattice deformation which increases the interlayer distance and raises the thickness of the thin MoS$_{2-x}$O$_x$ film \cite{Fleischauer_1987,Lince_1990,Buck_1991,Schmidt_1994,Dhall_2015} while forming oxide-containing patches on the surface. These fine oxide patches, spectroscopically determined to exist by Ko et al. \cite{Ko_2016}, can be seen in the scanning electron micrograph in \hyperref[fig:structural]{\textbf{Fig.2g}}. The contrast is due to the higher work function of MoO$_3$ (6.6 eV) \cite{Guo_2014} compared with that of MoS$_2$ ($\sim$4.04-4.47 eV) \cite{Lee_2016}. The structural modification undergone by the MoS$_2$ in the plasma chamber can also be linked to the change in its optical contrast over time (see \textbf{Supplementary Fig.10}). \newline
\indent In this sputtering-dominated regime, the surface roughness is seen to increase by over 1 nm at 28 seconds of exposure. However, the unchanging surface roughness up until 8 seconds indicates initial direct conversion of MoS$_2$ into a planar oxide. Most importantly, the edge height trend correlates with the electrical recovery discussed in the previous section, with a peak at 6 seconds. 
\onecolumngrid

{\setlength\intextsep{5pt}
\begin{figure}[H]
\centering
\includegraphics[scale=1]{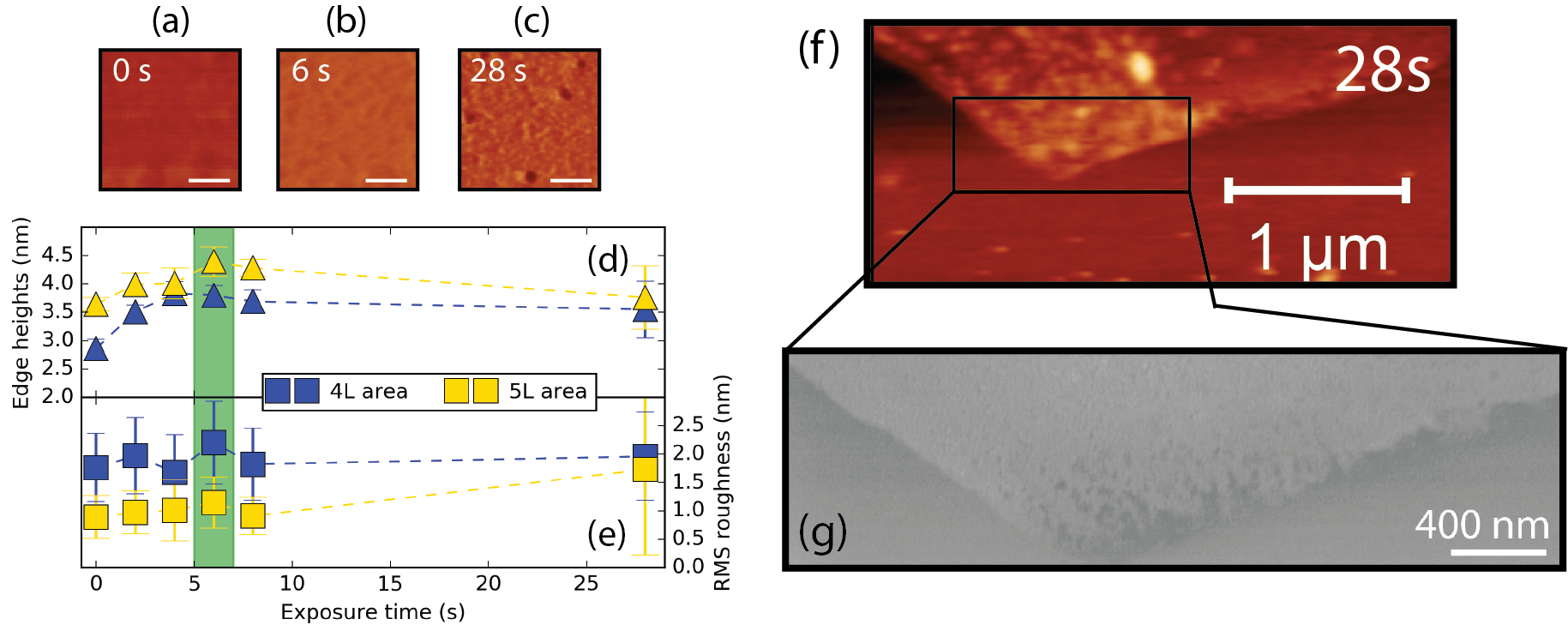}
\caption{{\footnotesize \textbf{(a)-(c) Phase maps of the same region of a 4L flake, showing notable corrugation of the surface as oxides are formed over time. Scale bars are all 200 nm. (d) Chart of edge heights extracted along line profiles after each exposure time (see all the raw height maps in \textbf{Supplementary Fig. 9}). The region in green is the edge height peak which correlates with the electrical recovery time at 6 s. (e) RMS surface roughness profiles extracted over time from height maps of the 4L and 5L regions of the flake. (f) AFM map of bottom edge of this flake after 28 s of plasma etching. Visible voids are seen along the bottom of the sample (g) SEM image of the corner of the same flake, exhibiting dark contrast pits on the edge, corresponding to oxidised MoS$_2$ regions.}}}
\label{fig:structural}
\end{figure}}
\twocolumngrid

\noindent \textbf{Spectroscopic analyses of the surface-bound oxide} \newline
\indent We investigate the change to the chemical content of our MoS$_2$ devices by employing Raman, photoluminescence (PL), energy dispersive X-ray (EDX) and X-ray photoelectron (XPS) spectroscopies. The Raman spectrum (\hyperref[fig:spectroscopies]{\textbf{Fig.3a}}) shows notable shifts in the characteristic peaks corresponding to the A$^{'}_{1}$ mode at $\sim$404 cm$^{-1}$ and the E$^{'}$ mode at $\sim$386.0 cm$^{-1}$ once the sample is exposed to the plasma. Accompanied by a 6-fold decrease in amplitude, the E$^{'}$ peak downshifts to $\sim$384 cm$^{-1}$ while the intensity of the A$^{'}_{1}$ peak decreases 5-fold and its position upshifts to $\sim$410 cm$^{-1}$, increasing the fitted peak separation from $\Delta \omega$ = 18 cm$^{-1}$ (characteristic for monolayer \cite{Li_2012}) to $\Delta \omega$ = 26 cm$^{-1}$ (see inset and fits in \textbf{Supplementary Fig.12}). The insertion of oxygen into the MoS$_2$ crystal lattice by the plasma can account for the change in dielectric screening environment and the restoring forces between adjacent MoS$_2$ molecular layers, thereby affecting the frequencies of both characteristic modes. An increase in $\Delta \omega$ occurs when MoO$_3$ replaces MoS$_2$ on the surface of the material \cite{Choudhary_2016,Ko_2016}. Conversely, the peak separation remains constant or is reduced when no oxides are detected after plasma treatment \cite{Kim_2016,Zhu_2016,Dhall_2015}. In addition, the asymmetric peak broadening of both peaks over time seen in \hyperref[fig:spectroscopies]{\textbf{Fig.3a}} has been associated with the presence of additional defect-induced phonons originating from oxide centres in plasma-treated MoS$_2$ \cite{Ko_2016}. \newline
\indent The PL intensity is reduced considerably, with a notable shift in peak positions, even just after 2 seconds of plasma exposure (\hyperref[fig:spectroscopies]{\textbf{Fig.3b}}). With further exposure, the emission is nearly fully quenched after 8 seconds. The bidirectional shifts of the A peak over time serve to illustrate band structure distortion induced by the plasma treatment. The photoluminescence quenching is due to the defect-induced midgap states that inhibit direct excitonic recombination \cite{Kang_2014, Choudhary_2016}. The associated quenching rate increases greatly with defect density, inhibiting the radiative recombination completely after 10 seconds of exposure (see \textbf{Supplementary Figs.13,14}). \newline
\indent Areal EDX mapping of the MoS$_2$ flakes \hyperref[fig:spectroscopies]{(\textbf{Fig.3c})} unveils the insertion of oxygen into the MoS$_2$ structure in a patch-like pattern, as has previously been suggested \cite{Islam_2014,neal2017p}. We note that plasma exposure time correlates with the gain in the oxygen K$_{\alpha}$ line relative to the sulfur K$_{\alpha}$ peak. We plot the relationship of the oxygen and sulfur peak intensities to the plasma exposure time (\hyperref[fig:spectroscopies]{\textbf{Fig.3d}}). Oxygen content has increased 3-fold in the first 10 seconds of exposure to the plasma. Importantly, areas high in O signal also show a reduced S signal, suggesting that the oxygen has replaced the sulfur in the MoS$_2$ lattice through an oxide-forming chemical reaction. 
\onecolumngrid

{\setlength\intextsep{5pt}
\begin{figure}[H]
\centering
\includegraphics[scale=1]{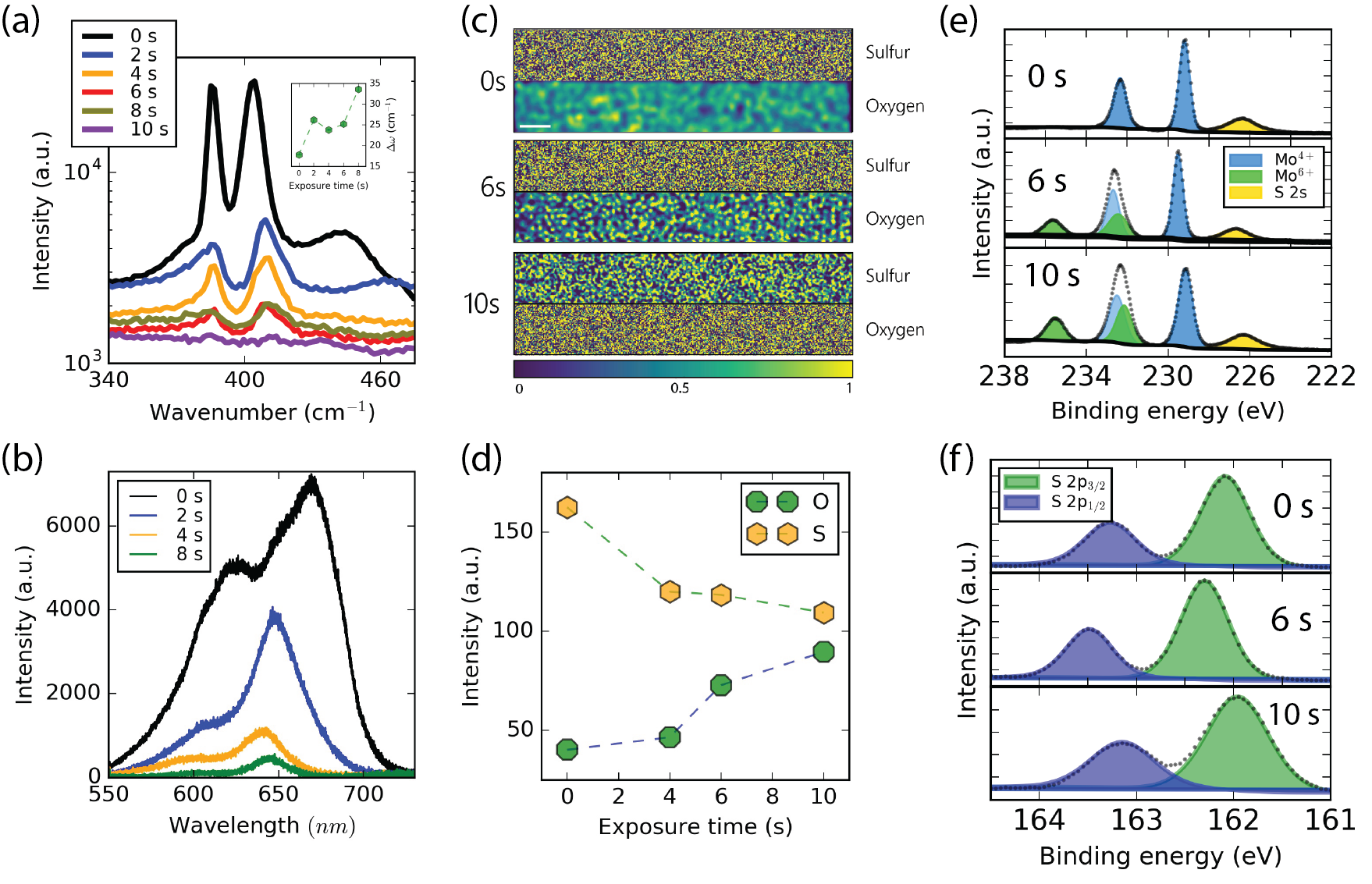}
\caption{{\footnotesize \textbf{(a) Raman spectrum of single layer MoS$_2$ over plasma treatment time shown on semi-log plot. The inset tracks the time evolution of the separation between the $E^{'}$ and $A^{'}_{1}$ peaks. (b) PL spectrum of monolayer MoS$_2$ as it changes over plasma exposure time. (c) Areal EDX mapping of sulfur and oxygen content in the sample at 0, 6 and 10 s of plasma treatment. The colorbar indicates normalised intensity of elemental signal collected at electron beam energy of 5 keV. Scale bar is 200 nm and the same for all maps. (d) Change of EDX intensity from oxygen and sulfur lines over exposure time in the areas shown in (c). (e) XPS spectra of the Mo 3d region showing increased MoO$_3$ content over time. (f) S 2p region of the XPS spectrum.}}}
\label{fig:spectroscopies}
\end{figure}}
\twocolumngrid

\indent \hyperref[fig:spectroscopies]{\textbf{Figure 3e}} shows XPS spectra of the Mo 3d region, indicating the increased presence of oxide species over exposure time. For the pristine sample, the peaks around 229 eV and 232 eV correspond to, respectively, the Mo$^{4+}$ 3d$_{5/2}$ and Mo$^{4+}$ 3d$_{3/2}$ spin-orbit split components. The 6 s spectrum shows a characteristic Mo$^{6+}$ 3d doublet attributed to MoO$_3$ \cite{scanlon2010}. After 10 s of exposure, the intensity of the trioxide-associated doublet increases further, with a significant ratio of the surface now containing MoO$_3$ (estimated at 30-40$\%$ from areas of each fitted component). In addition, a thickogram calculation \cite{cumpson2000thickogram} reveals that the intensity attenuation of the 10 s spectrum is consistent, within known parameters, with the presence of a bilayer of 2D-MoO$_3$, i.e. a bulk unit cell of $\alpha$-MoO$_3$ on the surface at this exposure time, and 61$\%$ of the bilayer of 2D-MoO$_3$ at 6 s (see details in \textbf{Supplementary Section 3}). \hyperref[fig:spectroscopies]{\textbf{Figure 3f}} demonstrates the S 2p region. Peak broadening is evident with increased plasma exposure time, indicating a change in chemical order of the surface. Sub-stoichiometric MoS$_{\ce{2-x}}$ has also been reported to cause the characteristic broadening of the S 2p doublet \cite{Baker_1999}, consistent with the picture of sulfur atoms being removed from the surface of the MoS$_2$ flakes.  \newline
\indent Most interestingly, both the Mo 3d and S 2p signals are upshifted after 6 s and downshifted after 10 s. It is widely accepted that MoO$_3$ can induce hole doping and concomitant downshifting of the MoS$_2$ peaks due to Fermi level realignment \cite{Zhu_2016,McDonnell_2014,Chuang_2014, neal2017p}. This is in agreement with our transfer curves, with significant threshold voltage shift to positive gate biases at higher plasma exposure times (\hyperref[fig:electrical]{\textbf{Fig.1c}}). The upshift at 6 seconds may correlate with the n-type doping observed in the transfer curves in \hyperref[fig:electrical]{\textbf{Fig.1b}}. \newline
\indent All the results demonstrate that the plasma-treated MoS$_2$ undergoes a continuing oxygen insertion and crystal structure distortion. However, plasma-generated MoO$_3$ is an insulator \cite{Islam_2014}. The electrical recovery at the 6 s mark indicates that an intermediary phase must exist between the pristine MoS$_2$ semiconductor and the MoO$_3$-rich insulator. This indicates that the 2D-MoO$_3$ phase at 6 s is markedly different from its bulk counterpart \cite{Zhang_2017}.    \\ \\
\noindent \textbf{Nanoscale effects of plasma etching at recovery time} \newline
\noindent Many etching mechanisms, some contradictory, have been proposed for the surface reaction of oxygen-containing plasma with MoS$_2$ \cite{Kim_2016,Yang_2013,Tao_2015,Ko_2016}. To uncover the nature of the elusive two-dimensional oxide phase, we go on to study the effects of the plasma etching on the nanoscale by aberration corrected scanning transmission electron microscopy (AC-STEM). \hyperref[fig:microscopy]{\textbf{Figure 4a}} shows a region of a bilayer MoS$_2$ flake after plasma treatment of 6 seconds. Notable damage occurs to the MoS$_2$ lattice at this exposure time. Regions of MoS$_2$ material are completely removed, in nanometre-sized regions. These pits deepen with increasing plasma dose and eventually become perforations. This etching phenomenon is seen to nucleate from individual defective sites, spanning only a few nanometres across. Some of these voids are missing a part of the top molecular layer of MoS$_2$ after 6 s, leaving behind a bare monolayer region underneath (as confirmed by simulation in \textbf{Supplementary Fig 28}). \newline
\indent Large-scale AC-STEM micrographs are presented in \textbf{Supplementary Fig.25}. These images were used to obtain statistics on the dimensions of voids formed by the plasma in the MoS$_2$ at the recovery time. \hyperref[fig:microscopy]{\textbf{Figure 4b}} demonstrates the distributions of the extracted widths and lengths of imaged voids on the bilayer flake. Yellow (green) histograms show the width (length) distributions. Length is here defined as the largest void dimension, while width is the dimension perpendicular to it. A positive correlation between the lateral dimensions of the etched voids is extracted from data fitting (see residuals in \textbf{Supplementary Fig.29}), showing the close-to-isotropic growth of the voids. The average area of a pore at 6 s is 12.5 $\pm$ 0.1 nm$^2$. At this time, the relative total percentage area covered by the voids from images sampled in the AC-STEM is $\sim$ 3.6$\%$.\newline
\indent Markedly, the nanoscale EDX and EELS mapping performed with the 60 pm electron beam probe return spectra suggesting very little oxygen presence (see \textbf{Supplementary Fig. 33}). It has been demonstrated that oxygen plasma interaction with molybdenum metal leads to the creation of volatile Mo oxides \cite{SABURI_2002}. We find that the plasma-formed oxide phase studied presently is volatile under ultra high vacuum. When the sample is left overnight in the in-situ testing system ($\sim$10$^{-9}$ mbar), the n-type depletion mode functionality is reversed by a drastic shift of V$_{\ce{TH}}$ towards positive gate biases (see \textbf{Supplementary Fig. 6}). Similarly, when inserted into the vacuum chamber of the AC-STEM overnight ($\sim$10$^{-9}$ mbar), the free-standing flakes lose their weakly-bound surface oxides. The inability to detect oxygen in these atomically-resolved voids leads us to infer that the oxide which was present initially and is responsible for the electrical recovery was sublimated at UHV conditions, leaving behind the underlying MoS$_2$ structure. 

\onecolumngrid

{\setlength\intextsep{5pt}
\begin{figure}[H]
\centering
\includegraphics[scale=1]{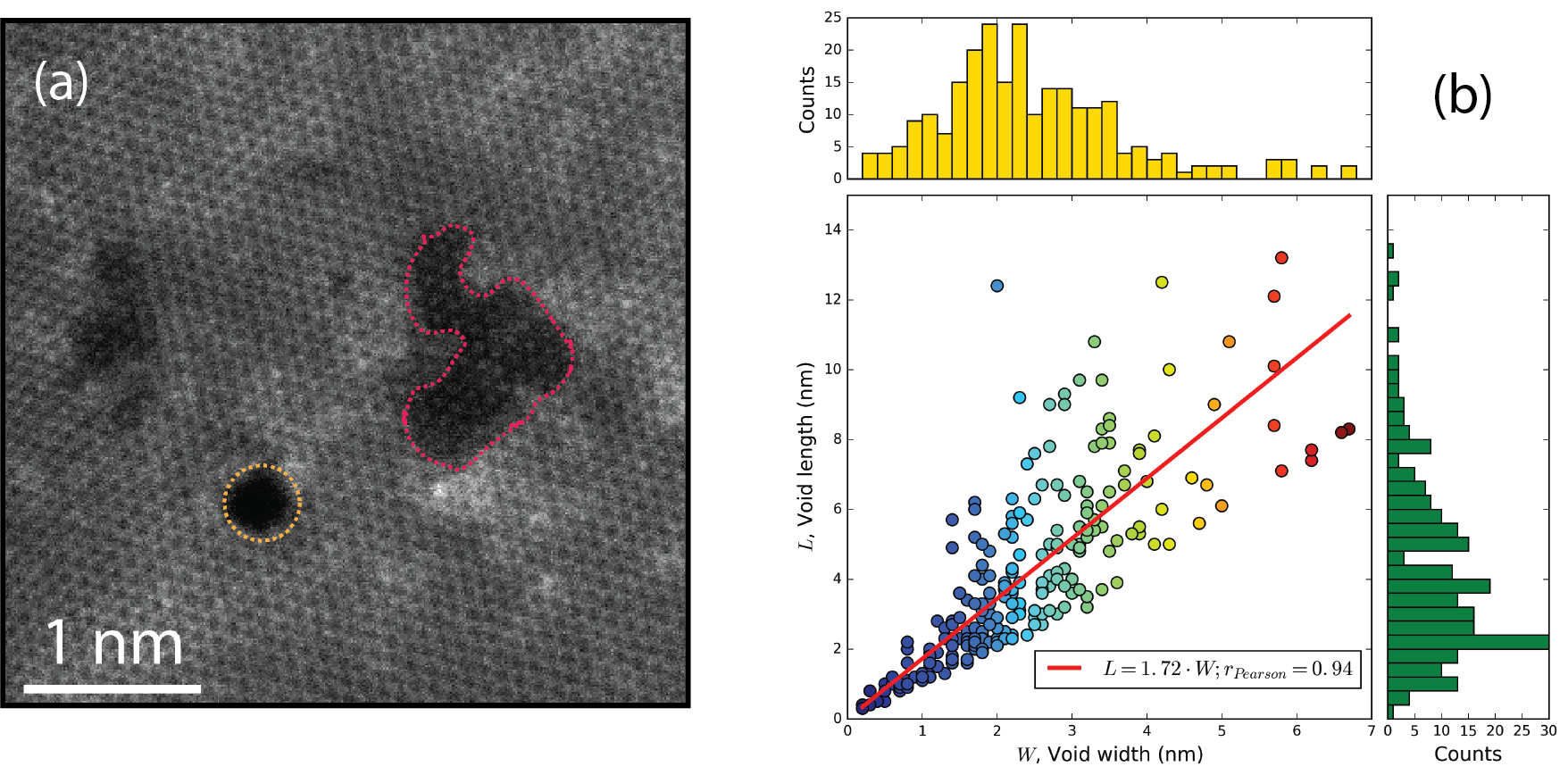}
\caption{{\footnotesize \textbf{(a) Typical AC-STEM micrograph showing nanoscale voids forming in bilayer MoS$_2$ after 6 s of plasma exposure. The region marked in red shows an etched pit with part of the top layer missing. The area marked in orange shows a perforation where no material remains. (b) Scatter plot and histograms visualising the distribution of the lateral dimensions of etched voids on this flake after 6 seconds of plasma exposure. The scatter data are color-mapped from cool to warm with increasing void width.}}}
\label{fig:microscopy}
\end{figure}}

\twocolumngrid

\onecolumngrid

{\setlength\intextsep{5pt}
\begin{figure}[H]
\centering
\includegraphics[scale=1]{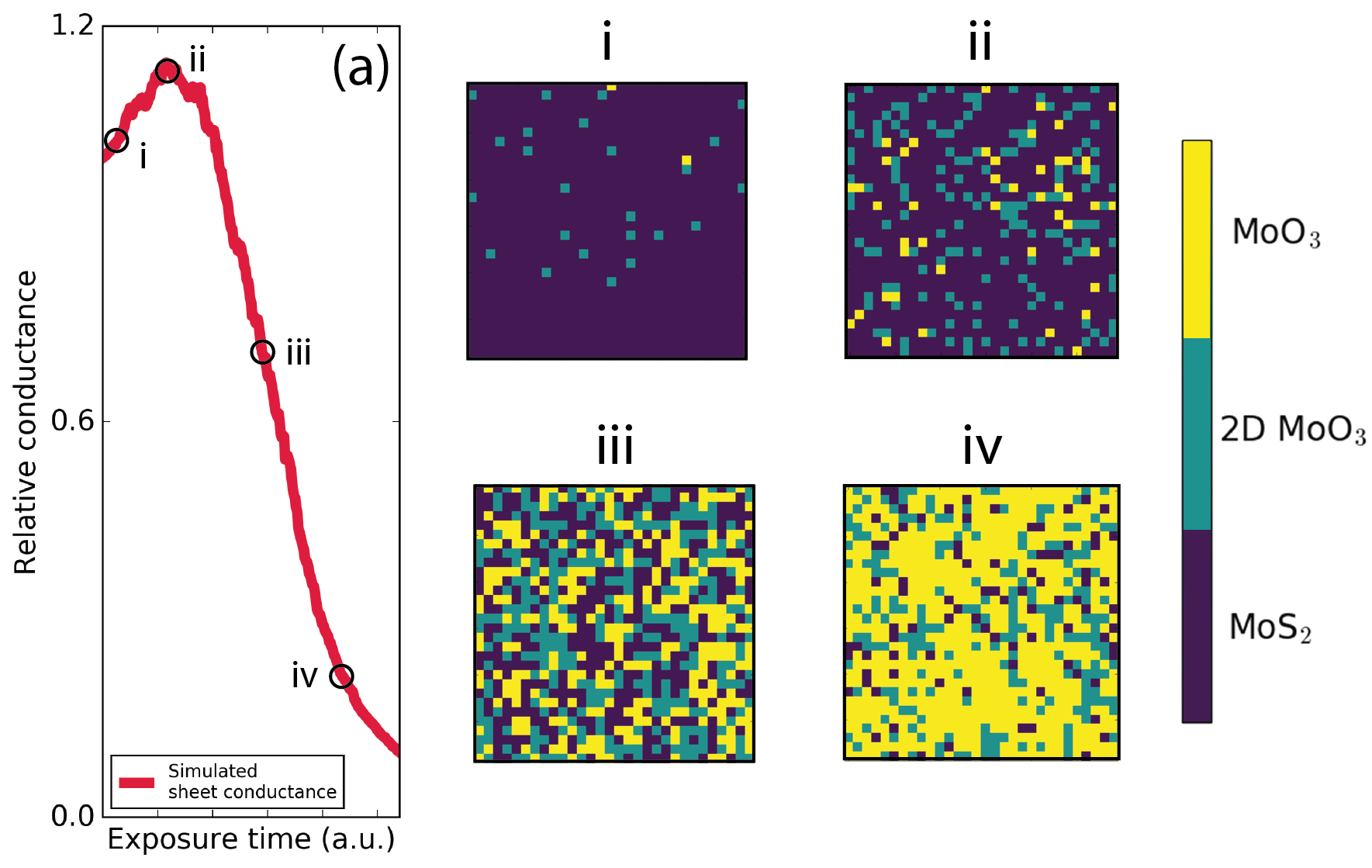}
\caption{{\footnotesize \textbf{Simulated sheet conductance of MoS$_2$ during the progressive conversion from MoS$_2$ to two-dimensional MoO$_3$ to MoO$_3$. Exposure time, which is in arbitrary units, charts the progress of the plasma-induced chemical oxidation as the simulation is iterated. One site undergoes transformation to the next phase during each iteration. Sites are not destroyed in this simulation, the plasma-etching process finishes when all sites reach the insulating MoO$_3$ phase. The distribution of phases in the conductive network at four different iteration stages is shown throughout i-iv, where the subsequent phases are color-mapped to the scale shown on the right. The lattice visualised in ii corresponds to the state of the network at the recovery time. The concentration of 2D MoO$_3$ at that stage is $\sim$ 15$\%$.}}}
\end{figure}
\label{fig:simulation}
\twocolumngrid

\noindent \textbf{Resistive network modelling of conductivity over time} \newline
\noindent 
Modelling of classical conductive networks has proven to be an excellent avenue to describe the global conductive properties of nanoscale devices based on local properties of a network \cite{Rocha_2015, OCallaghan_2016,Fairfield_2016}. An appropriate conductive network model was thus applied to approximate the effect of plasma-etching on the sheet conductance of MoS$_2$. \hyperref[fig:simulation]{\textbf{Figure 5a}} shows a plot of the relative conductance of a simulated MoS$_2$ network, whose nodes undergo conversion to highly conductive 2D-MoO$_3$ and insulating MoO$_3$ phases over time (in arbitrary units). The increase in sheet conductance yielded at the start of the simulation qualitatively mirrors the recovery peaks in \hyperref[fig:electrical]{\textbf{Fig. 1f}}. This can only be observed if the conductance of the 2D phase is much higher than that of MoS$_2$ and the relative transition rates between phases follow a relationship such that already defective sites are more likely to convert (see \textbf{Methods}). The relative concentration of conductors that exist in each phase is sampled throughout the simulation and visualised through the color-mapped lattices in i-iv. The sheet conductance rises initially as the concentration of 2D-MoO$_3$ increases. The conductance (and associated field effect mobility) reaches a peak, just as observed in experimental results in \hyperref[fig:electrical]{\textbf{Fig.1f}}, and proceeds  to drop off as conductors begin to transition to the insulating MoO$_3$.  \newline
\indent The differing transition rates affect not only the size of the conductive peak but phase concentrations that it occurs at (see \textbf{Supplementary Section 5}). The network map sampled at the conductance peak (ii) shows that the spike in sheet conductance occurs with a relative network coverage of 2D-MoO$_3$ at approximately 15$\%$. This differs from the experimentally determined areal void coverage of $\sim$ 4$\%$ at 6 s, but remains well under the site percolation threshold for a square lattice \cite{gebele1984site}. The difference may originate in the underestimation of the area of oxide-born defective sites from AC-STEM images of bilayer flakes. Some voids can become filled with adventitious hydrocarbons over time and/or be obscured by a top molecular layer depending on sample orientation when transferring to TEM grid. This resistive network model demonstrates qualitatively that a highly conductive intermediate oxide phase will serve to facilitate a recovery in the conductance and associated $\mu_{\ce{FE}}$ of a MoS$_2$ sheet over plasma exposure time.
\newline

\noindent \textbf{Outlook and conclusion} \newline
\noindent We have demonstrated a simple and reliable method to tune the electronic properties of few-layer MoS$_2$ FETs by using O$_2$:Ar (1:3) plasma. The apparent recovery of electrical conductivity is attributed to the temporary presence of a volatile two-dimensional phase of MoO$_3$, whose effect on the performance of the FET is self-limiting as further exposure results in physical etching and removal of the oxide. We have also inferred the existence of this 2D phase of MoO$_3$ from evidence collected from advanced spectroscopic and microscopic studies. Additionally, we have demonstrated with a robust simulation that the presence of a conductive phase on the surface of MoS$_2$ will induce a dose-dependent recovery in the conductance of the material network. Our results are of great importance to groups studying novel 2D TMDs and their low-dimensional oxides for future heterostructure van der Waals devices. 
\clearpage

\noindent \textbf{Methods} \\ \\
\textbf{MoS$_2$ exfoliation, identification and transfer:} MoS$_2$ flakes were mechanically exfoliated from commercially available
bulk molybdenite crystals (SPI Supplies) using the adhesive tape
method and deposited on a pre-cleaned
Si substrate capped by 285 nm of SiO$_2$. Samples of up to 10 layers in thickness were identified through optical contrast measurements and Raman spectroscopy. Electron
beam lithography was employed to define contacts, followed by deposition of metal film (5 nm Ti/35 nm Au) and lift off in acetone. Suspended MoS$_2$ samples were prepared utilising the stamp-transfer methodology \cite{Bie_2011} to move flakes from substrates onto TEM grids by etching away the SiO$_2$ surface underneath a polymer-embedded MoS$_2$ flake.

\noindent \textbf{Plasma treatment:} The on-chip MoS$_2$ FET devices were modified in a Fischione Instruments 1020
plasma cleaner, producing a 13.52 MHz field to ionise a 1:3 mixture of O$_2$:Ar
gas at a constant chamber pressure of $\sim5$ mbar. The samples were always exposed to the plasma for 2 seconds at a time, at the same position in the chamber (to within $\pm$ 1 mm), to control the accuracy of the experiment. After each exposure, the samples were removed and characterised electrically. 

\noindent \textbf{Electrical measurements:} The devices were globally
back-gated through the highly doped Si substrate and measured in a two-probe configuration at a pressure of 10$^{-4}$ mbar in the vacuum chamber of a scanning electron microscope. The source and drain terminals were provided by tungsten nanomanipulator
tips (Imina miBot) connecting the deposited contacts to an Agilent B2912A dual channel sourcemeter.

\noindent \textbf{Microscopy:} Transmission electron microscopy was carried out in an FEI Titan 80-300 operated at 300 keV, at a chamber
pressure of $4 \cdot 10^{-7}$ mbar. Atomic force microscopy was performed at ambient pressure in an Oxford Asylum system using cantilevers calibrated at 140 kHz. Aberration corrected scanning transmission electron microscopy was carried out in a NION UltraSTEM 200 system operated at 60 keV, at a chamber
pressure of $\sim 10^{-9}$ mbar.

\noindent \textbf{Spectroscopy:} As sampling efficiency from mechanically exfoliated flakes is extremely low, XPS was performed on larger flakes whose surface ($\sim2$ mm$^2$) was plasma-treated in the same way as all FET samples after deposition on Si/SiO$_2$ substrates. The system utilised a
monochromated Al K$_\alpha$ X-ray source with an Omicron EA 125 hemispherical
analyser set to a pass energy of 19 eV, giving a combined instrumental and source resolution of 0.50 eV. The spectra for these samples were fitted with 2H polytype peaks, as is usual for mechanically exfoliated MoS$_2$ flakes. PL spectroscopy was performed on substrate-supported flakes using an excitation wavelength of 405 nm. Raman spectroscopy was carried out at atmospheric pressure with a Horiba Jobin-Yvon 488 nm laser equipped with 1200 grooves/mm and a CCD camera. Acquisition time was fixed at 10 acquisitions per second. A 100 $\times$ objective lens was used. The laser spot size was $\sim$1 $\mu$m, while the power of the laser was kept below 1 mW. EDX mapping was done on suspended samples using a Bruker
Nano XFlash 5030 detector in a Zeiss Supra SEM at 5 keV, with a step size of
0.7nm/px. \newline
\noindent \textbf{Simulation of conductive networks:} The computational model begins with a resistive network of identical conductors of magnitude $g_{\ce{S_2}} = 1$. During each iteration of the simulation, one random conductor transitions from its current phase to the next phase with a certain probability, unless that conductor is already in the final MoO$_3$ phase. If it does not transition then one of the adjacent sites is chosen and the transition check process is repeated. The probabilities represent the differing transition rates that occur between phases. While the transition rates are experimentally unknown, the assumption is made that they progress such that MoS$_2$ $\overset{p_1}{\rightarrow}$ 2D-MoO$_3$ $\overset{p_2}{\rightarrow}$ MoO$_3$, where p$_1$, p$_2$ indicate relative conversion probabilities for each process and p$_2$ $ > $ $p_1$. This relationship stems from the fact that the MoS$_2$ basal plane is chemically unreactive, but any defective nucleation sites will be more likely to facilitate chemical reactions once they are formed \cite{kc2015surface}. The sheet conductance is then calculated using Kirchhoff's and Ohm's laws (see \textbf{Supplementary Section 5}). The iterations are continued until all conductors are in the insulating MoO$_3$ phase. Iterations in the simulation are a proxy to the plasma-exposure time, with a certain number of phase transitions (or iterations) per unit time. \newline

\section*{Acknowledgements}

We are grateful to members of staff at the Advanced Microscopy Laboratory, CRANN, Trinity College Dublin for their continued technical support. We thank C. P. Cullen and S. Callaghan for fruitful discussions regarding XPS. The work at the School of Physics and the Centre for Research on Adaptive Nanostructures and Nanodevices at Trinity College Dublin is supported by Science Foundation Ireland [grant No: 11/PI/1105, 12/TIDA/I2433 07/SK/I1220a and 08/CE/I1432] and the Irish Research Council [grant No: GOIPG/2014/972 and EPSPG/2011/239]. 

\section*{Author contributions}

J.J analysed the data, created figures and wrote the manuscript with input from C.O'C and H.Z. I. O'R. discovered the phenomenon. J.J and Y.Z. conducted subsequent plasma exposures and electrical tests. C.O'C. and E.W. carried out the resistor network modelling. Y.Z. performed Raman experiments. P.M. and J.J. carried out SEM imaging and EDX measurements. D.F. and J.J. performed HRTEM and STEM. C.D. acquired AC-STEM images and associated EELS and EDX maps. A.S analysed the AC-STEM results and performed QSTEM simulations. J.J., Y.Z. and J.G. carried out the PL measurements. C. McG. and J.J. performed XPS experiments and analysis. D.K. and J.J. carried out AFM experiments. M.S.F., A.L.B., J.J.B. and V.N. oversaw the experimental work. H.Z. conceived the study and supervised the project. All authors have given approval for the final version of the manuscript.

\section*{Additional information}

\noindent Supplementary information containing additional experimental data for each of the sections discussed in this manuscript is available online. \\
\textbf{Data availability}: Correspondence and requests for materials should be addressed to H.Z. Raw data (code and source data for graphs) generated and/or analysed during the current study are available in the Zenodo repository at: \href{https://www.zenodo.org/record/809442}{DOI:10.5281/zenodo.809442}

\section*{Competing financial interests}
The authors declare no competing financial interests.

\end{document}